\documentclass[aps,twocolumn,prl,showpacs]{revtex4}
\usepackage{amssymb,amsbsy,graphicx,subfigure,times,bbm}
\usepackage{hyperref}
\usepackage[latin1]{inputenc}
\usepackage{amsmath,amsfonts,amssymb,graphics,graphicx,epsfig,color}
\usepackage{subfigure}
\usepackage{fancyhdr,makeidx}
\usepackage{color}

\newcommand{\R}{\mathbbm{R}}

\newcommand{\id}{\mathbbm{1}}

\newcommand{\tr}{{\rm Tr}\,}

\newcommand{\gr}[1]{\boldsymbol{#1}}
\newcommand{\be}{\begin{equation}}
\newcommand{\ee}{\end{equation}}
\newcommand{\bea}{\begin{eqnarray}}
\newcommand{\eea}{\end{eqnarray}}
\newcommand{\ket}[1]{|#1\rangle}
\newcommand{\bra}[1]{\langle#1|}

\newcommand{\sig}{\gr{\sigma}}

\newcommand{\eq}[1]{Eq.~(\ref{#1})}

\begin{document}
\title{Teleportation fidelities of squeezed states from thermodynamical 
state space measures}
\author{Alessio Serafini, Oscar C.O. Dahlsten, and Martin B. Plenio}
\affiliation{Institute for Mathematical Sciences, 53 Prince's Gate, Imperial College London, London SW7 2PG, UK\\
and QOLS, Blackett Laboratory, Imperial College London, London SW7 2BW, UK}
 
\begin{abstract}
{We introduce a ``microcanonical'' measure (complying with the `general canonical principle')
over the second moments of pure bosonic Gaussian states under an energy constraint.
We determine the average fidelity for the teleportation of states 
distributed according to such a measure and compare it to a threshold obtained from a feasible classical strategy.
Furthermore, 
we show that, under the proposed measure, the distribution of the entanglement concentrates 
around a finite value 
at the thermodynamical limit and, in general, the typical entanglement of Gaussian states with 
maximal energy $E$ is {\em not} close to the maximum allowed by $E$.}
\end{abstract}
\maketitle

\noindent{\bf \em Introduction} -- Besides having been at the core of theoretical and experimental quantum optics
right from its early stages,
Gaussian states have recently acquired a major role
in quantum information science, 
in the so called `continuous variable' (CV) scenario \cite{rev}.
Indeed, some of the most spectacular implementations 
of quantum information protocols to date are based on Gaussian states,
with the prominent example of deterministic teleportation \cite{furusawa98}.
In the analysis of most such implementations, 
the proper assessment of figures of merit requires the average over a distribution (a ``measure'') of states 
in which input quantum information is encoded.
For instance, in the case of quantum teleportation of coherent states, 
the theoretical average fidelity (between input and output states) 
is determined by assuming a particular distribution of input coherent states \cite{braunstein00}.
In the present paper we propose a measure on the set of pure Gaussian states,
 whose introduction will be thoroughly motivated by fundamental statistical arguments \cite{popescu05}. 
We shall focus on the second moments of the quadrature operators 
(while the measure usually employed to analyze teleportation of coherent states \cite{braunstein00} 
essentially encompasses first moments),
covering the whole set of pure Gaussian states with null first moments. 
As we will mention later,
first moments may be accomodated as additional variables in the presented framework.

The importance of determining a suitable measure over a set of states
is not merely a theoretical issue, as the evaluation 
of classical thresholds for the figures of merit is crucial in establishing whether 
practical realizations of quantum protocols actually out-perform competing classical strategies \cite{braunstein00}.
We shall thus apply the proposed measure to determine 
the average teleportation fidelity of pure Gaussian states 
with varying second moments, and shall compare such a fidelity
to a corresponding ``classical'' threshold.
Moreover, to further illustrate the potentialities of a measure on second moments,
we will address the ``typical'' entanglement \cite{typical} 
of pure Gaussian states under an energy constraint. 
The very construction of the measure will imply that
the distribution of the von Neumann entropy 
of any finite subsystem `concentrates', both at the thermodynamical limit and for finite 
numbers of modes, around a finite `thermal' average, 
well away from the allowed maximum.

\noindent{\bf \em Preliminary facts and notation} --
We consider continuous variable quantum mechanical systems 
described by $n$ pairs of canonically conjugated 
operators $\{\hat x_j,\hat p_j\}$ with continuous spectra.
Grouping the canonical operators together in the 
vector $\hat R=(\hat{x}_1,\ldots,\hat{x}_n,\hat{p}_1,\ldots,\hat{p}_n)^{\sf T}$ allows one to 
express the canonical commutation relations as
$
[\hat R_j,\hat R_k] = 2i\,\Omega_{jk} 
$, 
where the `symplectic form' $\Omega$ has entries 
$\Omega_{jk} \equiv \delta_{j+n,k}-\delta_{j,k+n}$ for $j,k=1,\ldots,2n$.
Any state of an $n$-mode CV system is described by a positive, trace-class 
operator $\varrho$. 
For any state $\varrho$,
let us define the $2n\times 2n$ matrix of second moments, 
or ``covariance matrix'' (CM), 
${\gr\sigma}$ with entries 
${\sigma}_{jk} \equiv \tr{[\{\hat R_j , \hat R_k\} \varrho]}/2
-\tr{[\hat R_j \varrho]}\tr{[\hat R_k \varrho]}$.
In the following, we will refer to the `energy' of a state $\varrho$ as to the 
expectation value of the operator $\hat H_0 = \sum_{j=1}^{n} (\hat{x}_j^2+\hat{p}_j^2)$
(note that, in our convention, the vacuum of a single mode has energy $2$).
This definition corresponds to the energy of a free electromagnetic field in the 
optical scenario (and to decoupled oscillators in the general case). 
Neglecting first moments, 
the energy is determined by the second moments according to 
$
\tr{(\varrho\hat{H}_0)} = \tr(\sig)
$.

Gaussian states are defined as the states with Gaussian characteristic functions 
and quasi-probability distributions.
All pure Gaussian states can be 
obtained by transforming the vacuum under unitary operations generated 
by polynomials of the second order in the canonical operators. 
Operations generated by first order polynomials in the quadratures
correspond to local displacements in the first moments, and will thus be disregarded. 
As for second order transformations, they can be mapped 
into the group $Sp_{2n,\R}$ of real {\em symplectic transformations}, by virtue of the so called 
{\em metaplectic representation} 
(recall that $S\in SL({2n,\R}) \, :\;
S\in Sp_{2n,\R} \Leftrightarrow S^{\sf T} \Omega S = \Omega$). 
As a consequence of such a mapping, the CM $\sig$ of any pure Gaussian 
state can be written as $\sig=S^{\sf T}S$ \cite{rev,nosotros}.

\noindent{\bf \em General canonical principle and microcanonical measure.} -- 
We will now proceed to define a measure over the set of pure Gaussian states, 
which will be referred to as `microcanonical' (for reasons which will be clear shortly).
Henceforth, the shorthand notation $\overline{x}$ will stand for the average of the quantity $x$ 
with respect to such a measure. Since we will adopt a constructive approach, based on the gradual enforcement of specific conditions on the measure, 
the notation $\overline{x}$ will appear, with no ambiguity, before the definition of the measure itself.

Because the symplectic group is non-compact, an invariant Haar measure 
on the whole group (from which a measure for the second moments of 
pure Gaussian states could be derived via the equation $\sig=S^{\sf T}S$) 
would be non-normalizable (and thus
``unphysical'', giving rise to distributions with unbounded statistical moments).
The first natural prescription to tame the non-compact nature of the group consists in the introduction 
of a constraint on the total energy of the system, which we will denote by $E$ 
(hence the designation ``microcanonical'' attached to the measure). 
Even so, no natural invariant measure emerges.
However, let us recall that an arbitrary symplectic transformation $S$ can be decomposed as 
$
S = O' (Z\oplus Z^{-1}) O
$, 
where $O, O' \in K(n)\equiv Sp(2n,\R)\cap SO(2n)$ are orthogonal
symplectic transformations, while
$Z$ is an $n\times n$ diagonal matrix 
with eigenvalues $z_{j}\ge 1$ for $1\le j\le n$ \cite{pramana}.
The set of transformations of the form $Z\oplus Z^{-1}$ 
is a non-compact subgroup of $Sp_{2n,\R}$ (corresponding to local squeezings). 
The virtue of such a decomposition, known as  the `Euler' decomposition,
is immediately apparent, as it allows  one to distinguish between 
the degrees of freedom of the compact subgroup 
(`angles', collectively denoted by $\vartheta$, which do not affect the energy) 
and the degrees of freedom $z_j$'s with non-compact domain. 
In particular, applying  the Euler decomposition to the CM $\sig$ of generic pure states leads to
$
\sig = O^{\sf T} (Z^2\oplus Z^{-2}) O 
$.
Moreover, we recall that the compact subgroup $K(n)$ is isomorphic to $U(n)$ 
\cite{pramana}.
As dictated by the Euler decomposition, we assume
the $n^2$ parametres $\vartheta$ of the transformation $O\in K(n)$ to be distributed according 
to the Haar measure of the compact subgroup $K(n)$,  
which can be carried over from $U(n)$ through the isomorphism recalled above
and whose infinitesimal element will be denoted by ${\rm d}\mu_{H}(\vartheta)$.

We are thus left with the parameters $z_{j}$'s alone, for which a `natural' measure 
has not yet emerged. To constrain the choice of such a measure, 
we will invoke a fundamental statistical argument. In their alternative, `kinematical' 
approach to statistical mechanics, Popescu {\em et al.}~\cite{popescu05} define a general principle, 
which they refer to as {\em general canonical principle}, stating that
``Given a sufficiently small subsystem of the universe, almost every pure state of the universe 
is such that the subsystem is approximately in the `canonical state' $\varrho_{c}$.''
The `canonical' state $\varrho_c$ is, in our case, 
a Gaussian thermal state, with CM $\sig_c=(1+T/2)\id$.
Here the `temperature' 
$T$ is defined by passage to the thermodynamical limit, that is  
for ${n\rightarrow\infty}$ and ${E\rightarrow\infty}$,   
$(E-2n)/n\rightarrow T$ (assuming $k_{B}=1$ for the Boltzmann constant). 
For ease of notation, in the following,
the symbol $\simeq$ will imply that the equality holds at the thermodynamical limit, 
{\em e.g.}:~$(E-2n)/n\simeq T$.
Because the state $\varrho_c$ is Gaussian with null first moments, the general canonical principle
can be fully incorporated into our restricted (Gaussian) setting. 
To this aim, let us recast the principle in terms of 
mathematical conditions to be fulfilled by the underlying measure on pure Gaussian states. 
Recall that partial tracing (obviously a Gaussian operation) amounts, at the level of CM's, 
to simply pinching the submatrix of $\sig$ pertaining to the relevant modes.
Then, in order for the measure 
to comply with the general canonical principle, one has to require
\be
\overline{\sigma_{jk}} \simeq (1+T/2) \delta_{jk} \, , \quad 
\overline{\sigma_{jk}^2} \simeq (1+T/2)^2 \delta_{jk} \; . \label{condave} 
\ee
The second condition rephrases the prescription 
``almost every pure state'', requiring `concentration of measure' at the thermodynamical 
limit (in fact, in conjunction with the first equation, 
it implies that the variance of the entries of the CM vanishes at the thermodynamical  limit). 
The previous conditions, which are highly desirable to single out a measure naturally endowed 
with physical and statistical significance, will greatly restrict the possible choices for the 
distribution of the variables $z_j$'s. 

In order to show this, we first work out the averages of the entries of $\sig$
over the Haar measure of the compact subgroup. 
This task can be accomplished relying only on some basic properties of the integration 
over the unitary group, derived from simple symmetry arguments 
(see \cite{aubertlam}, a detailed derivation can be found in \cite{nosotros}),
leading to 
\be
\overline{\sigma_{jk}} = \frac{1}{2n}\Big[\sum_{l=1}^{n} \overline{(z^{2}_{l}+z_{l}^{-2})}\Big]\delta_{jk}
\equiv
\frac{1}{2n}\Big[\sum_{l=1}^{n} \overline{E_{l}}\Big]\delta_{jk} \, . \label{ave}
\ee
The convenience of a parametrisation through the variables $E_{j}\equiv(z^2_j+z^{-2}_j)$, 
representing the local energies of the decoupled modes, is now apparent.
The same arguments, based on symmetry and normalization (reflecting 
orthogonality), can be applied to the average $\overline{\sigma_{jk}^2}$, leading to
\be
\overline{\sigma_{jk}^2} = \frac{1}{4n^2}
\Big[\sum_{l_1,l_2=1}^{n} \overline{E_{l_1}E_{l_2}}\Big] \delta_{jk} \, . \label{var}
\ee
Now, the desired agreement of Eqs.~(\ref{ave},\ref{var}) with Eq.~(\ref{condave}) 
single out a restricted class of measures 
for the variables $E_{j}$'s. 
Most notably, 
{\em any measure such that the local energies $\{E_{j}\}$ are, {\em at the thermodynamical limit}, 
independent, identically distributed (``i.i.d.'') variables with average $\overline{E}\equiv(T+2)$ 
complies with the general canonical principle} \cite{gennote}. 
In point of fact, at the thermodynamical  limit, 
only the averages $\overline{E_j E_k}$ with $j\neq k$ 
matter in the computation of the variance, as their number scales as $n^2$, 
while the number of terms in $\overline{E_j^2}$ is clearly linear in $n$, 
and their contribution gets suppressed by the factor $1/n^2$ (deriving from the integration 
over the Haar measure of a term of degree two in the compact transformations' entries). 
The same argument holds for the square of the quantity $\overline{\sigma_{jk}}$ of \eq{ave}.
For i.i.d. variables, $\overline{E_{j}}\simeq \overline{E}$ and 
$\overline{E_{j}E_{k}}\simeq \overline{E}^2$ $\forall\,j\neq k$, thus implying the vanishing 
of the variance at the thermodynamical  limit.

To complete the definition of the measure, we have to specify a distribution of the $E_{j}$'s 
in agreement with the previous requirements.
Recovering the energy constraint $E$, 
we will assume a Lebesgue (`flat') measure for the local energies $E_j$'s 
inside the region $\Gamma_{E}=\{\gr{E} : |\gr{E}|\le E\}$, 
bounded by the linear hypersurface of total energy $E$ (here, ${\gr E}=(E_1,\ldots,E_n)$ 
denotes the vector of energies, while $|\gr{E}|=\sum_{j=1}^{n}E_j$). 
More explicitly, denoting by ${\rm d}\,p(\gr{E})$ the probability of the occurrence of the energies $\gr{E}$ 
and by ${\cal V}=(E-2n)^{n}/n!$ the volume of the region $\Gamma_E$, one has 
$
{\rm d}\,p(\gr{E}) =  {\rm d}^n\gr{E} /{\cal V} \equiv ({\rm d}\,E_1\ldots{\rm d}\,E_n)/{\cal V}$ if $\quad \gr{E}\in \Gamma_{E}$
and ${\rm d}\,p(\gr{E}) = 0$ otherwise. 
Notice that such a flat distribution is the one maximising the entropy in the knowledge of the local energies 
of the decoupled modes. In this specific sense such variables have been privileged, on the basis
of both mathematical (Euler decomposition and Haar averaging over the compact subgroup) and physical 
(general canonical principle and analogy with the microcanonical ensemble) grounds.
In full analogy with the equivalence of statistical ensembles, 
the $E_{j}$'s become i.i.d. at the thermodynamical limit. 
In fact, the marginal density of probability $P_{n}(E_j,E)$ for each of the energies $E_j$ given by
$
P_{n} (E_j,E) = \frac{n}{E-2n} \left( 1- \frac{E_j-2}{E-2n} \right)^{n-1} 
$.
At the thermodynamical limit, 
the upper integration extremum for each $E_j$ diverges and 
$P_{n} (E_j,E)\rightarrow {\rm e}^{-\frac{E_j-2}{T}}/T$, 
so that the decoupled energies are distributed according to 
independent Boltzmann distributions with the  parameter $T$ playing the role of a temperature,
and their averages satisfy \eq{condave}. 
The microcanonical measure is thus consistent with the general canonical principle \cite{1stmoments}. 
The `microcanonical' average $\overline{Q}$ over pure Gaussian states at energy $E$ of 
the quantity $Q(\gr{E},\vartheta)$ determined by the second moments alone 
will thus be defined as 
$
\overline{Q} = {\cal N}\int {\rm d}\mu_{H}(\vartheta) \int_{\Gamma_{E}} {\rm d}\gr{E} Q(\gr{E},\vartheta)
$, 
where the integration over the Haar measure is understood to be carried out over the whole compact domain 
of the variables $\vartheta$.

\noindent{\bf \em Typical entanglement} -- Here, we concisely address the statistical properties 
of the bipartite entanglement of pure Gaussian states under the microcanonical measure \cite{nosotros}.
Let us consider the von Neumann entropy $S$ of the reduced state $\varrho_m$ 
of a finite number of modes $m$ with CM $\gr{\gamma}$, 
thus quantifying its entanglement with the remaining $(n-m)$ modes of the globally pure state.
At the thermodynamical limit the distribution of the CM $\gr{\gamma}$ 
concentrates, {\em with vanishing variance}, around a thermal state with CM $(T/2+1)\id$, 
according to the general canonical principle.
Therefore, the distribution of the 
von Neumann entropy of the reduction $S$ concentrates around the von Neumann entropy of a
thermal state \cite{nosotros}. In formulae: $\overline{S}\simeq m f(1+T/2)$ 
and $(\overline{S^2}-\overline{S}^2)\simeq 0$, where 
$f(x)\equiv(x+1)\log_{2}[(x+1)/2]/2-(x-1)\log_{2}[(x-1)/2]/2$. Notice that the maximal 
local von Neumann entropy of any (finite or infinite) subsystem diverges 
at the thermodynamical limit (as, in principle, all the infinite energy could
be concentrated in only two modes -- owned by the two distinct subsystems --, 
thus yielding an infinite entropy for each subsystem).
For finite $n$, 
the microcanonical measure is apt to be investigated numerically 
by direct sampling, allowing one to  
study the distribution of the von Neumann entropy for different $m$, $n$ and $E$.  
Even for small $n$ -- well before the onset of thermodynamical concentration of measure around 
the finite thermal average -- the entanglement of pure Gaussian states distributes,
for small enough energies, 
around values generally distant from the finite allowed maximum ({\em
e.g.}, for $m=1$ and $E= 10n$, the difference between the maximum
and the average $\overline{S}$ is, respectively, 
$4.0$ and $13.6$ standard deviations for $n=5$ and $n=20$). 
This is at striking variance with results obtained, adopting different measures, 
in finite dimensional systems \cite{typical}. 
The equipartition of energy, imposed by the general canonical principle, prevents the entanglement 
of finite subsystems from concentrating around the maximum.
Notice that the concentration of measure would also occur for a distribution of states with {\em fixed} 
total energy $E$ (and all the variables $\{E_j\}$ still Lebesgue-distributed under such a constraint). 
This is the case as such a measure is equivalent, at the thermodynamical limit, to the 
one we are considering (they both converge to exactly the same Boltzmann distribution).

\begin{figure}[t!]
\begin{center}
\includegraphics[scale=1]{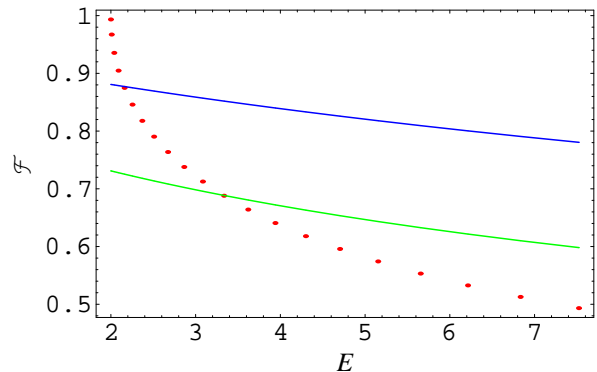}
\caption{Average teleportation fidelity for $r=1$ (blue curve) and $r=0.5$ (green curve) 
versus the heterodyne threshold (red dots), as functions of the maximal energy $E$. \label{classfido}}
\end{center}
\end{figure}

\noindent{\bf \em Teleportation fidelities} -- Let us now consider a practical situation, in which 
two parties (Alice and Bob) want to communicate through quantum teleportation. 
Instead of contenting themselves with coherent states, Alice and Bob are interested in exchanging
single-mode pure Gaussian states with arbitrary CM (wherein the quantum information 
is encoded). To this aim, they employ the usual CV teleportation scheme, based on homodyne 
measurements and on the sharing of a two-mode squeezed state 
with squeezing parameter $r$ (essentially quantifying the entanglement exploited in the 
teleportation process, see \cite{rev} for a detailed description of the scheme). 
Suppose, quite reasonably, that Alice generates (and sends to Bob) states 
with a flatly distributed energy up to a maximal value $E$ and random optical phase,
whose distribution is thus described by the microcanonical measure.
The microcanonical average fidelity $\overline{\cal F}$ 
(defined as the average, over the distribution of input states and of measurement outcomes, 
of the overlap $|\bra{in}{out}\rangle|^2$
between the input state $\ket{in}$ and output state $\ket{out}$)
can be straightforwardly determined \cite{pirandola}, and found to be 
$\overline{\cal F} = 2\,{\rm e}^{2r}(\sqrt{1+\,{\rm e}^{-4r}+E\,{\rm e}^{-2r}}-1-\,{\rm e}^{-2r})/(E-2)$.
To properly assess the effectiveness of the standard teleportation protocol in transmitting second moments, 
let us compare the previously obtained fidelity to an appropriate 
``classical threshold'' ${\cal F}_{cl}$ (as customary in the 
literature on teleportation, ``classical'' refers to a procedure where no entanglement is exploited). 
The kind of classical strategy we will consider is as follows: Alice measures her mode by heterodyne detection 
(corresponding to the positive operator valued measurement with elements $\ket{\alpha}\bra{\alpha}/\pi$, 
where $\ket{\alpha}$ is a coherent state \cite{parisbook}) 
and sends her result (a complex number $\alpha$, representing the heterodyne signal) 
to Bob who reproduces a centered, pure state (belonging to the original distribution) \cite{classinote}.
Bob's choice,  
depending on $\alpha$, has been optimised numerically and the resulting fidelity 
has been averaged over the input distribution of states, to obtain the ``heterodyne''
classical threshold $\overline{{\cal F}_{cl}}$ (lower bounding 
the actual classical threshold).
For sufficiently small $E$, one can also approximate such a threshold 
-- obtained numerically -- as 
$\overline{{\cal F}_{cl}}\approx 1-k\,{\rm arcsinh}(\sqrt{E-2})$, for $k=0.317576$ 
(such a fit is reliable within $0.002$ in the range $2\le E\le 8$).
Comparing the experimental average fidelity to the
previous formula would tell Alice and Bob whether 
their precious quantum entanglement is offering an actual advantage over
a viable, `cheaper' protocol based on disjoint measurements and reconstructions
of the states \cite{pro}.
Fig.~\ref{classfido} shows that, for a given $r$, the classical strategy beats CV teleportation 
for small enough $E$. Actually, this fact is not surprising: it simply results from the inadequacy of the 
standard teleportation protocols when the input alphabet is overly restricted, and 
occurs in the teleportation of coherent states as well, 
if the choices of the coherent amplitudes are sufficiently constrained (the reader might think about 
the limiting instance for which the vacuum is the only input state: then 
the teleportation protocol, completely based on probabilistic measurement outcomes, fails to yield
a fidelity equal to one, whereas the classical protocol is set to always return the vacuum in such an 
instance).
However, there always exists a value of $E$ above which the CV protocol starts outperforming 
the classical strategy. For instance, in the experimentally realistic case $r\approx1$, such a threshold
is remarkably low, being around $E\approx2.16$. 
On the other hand, for any value of $E$, the teleportation protocol may always exceed the 
classical threshold for high enough $r$. Clearly, for any finite $E$, 
one has $\lim_{r\rightarrow\infty}\overline{{\cal F}}=1$ (teleportation with unlimited 
resources is always perfect). 
Also, numerics unambiguously show that 
$\lim_{E\rightarrow\infty}\overline{{\cal F}_{cl}}=0$.
Likewise however, for any finite $r$, one has $\lim_{E\rightarrow\infty}\overline{{\cal F}}=0$. 
This limiting behaviour is quite remarkable and might inspire future inspection into the matter: 
when the alphabet of states is enlarged to encompass all the possible second moments, the fidelity 
of the standard teleportation protocol vanishes (as opposed to what happens for coherent states, where 
the fidelity stays constant even if the alphabet is extended over an unbounded domain). 
This suggests that, possibly, a modified protocol where Bob can act unitarily 
on the second moments could grant better fidelities when the teleportation of second moments 
is concerned.

\noindent{\bf \em Outlook} -- The study of generic entanglement 
and of figures of merit for teleportation
are only examples of the potential applications of the 
microcanonical measure. 
For instance, the compliance with the general canonical 
principle renders the measure suitable to describe the thermalization of systems 
in dynamical situations \cite{gemmer01}. Relating the measure to distributions derived from a 
randomizing process (in the spirit of Ref.~ \cite{oliveira})
is a further line of development opened up by the present investigation.

\noindent Helpful discussions with G.~Adesso, D.~Gross, J.~Eisert and B.~Collins 
are acknowledged. 
A.~S.~was funded by a Marie Curie Fellowship, 
O.~D.~by the Institute for Mathematical Sciences 
of Imperial College, 
M.B.~P.~was funded by the EPSRC QIPIRC,
The Leverhulme Trust, EU Integrated Project QAP,
and the Royal Society.


\begin{thebibliography}{99}
\bibitem{gennote}Even though the agreement between 
the two pairs of equations could allow for a more general choice,
i.i.d. variables are by far the most convenient assumption 
for the sake of clarity of both mathematical description 
and physical interpretation.
\bibitem{1stmoments}The first moments $R$ (here $R$ is a $2n$-dimensional real vector, with 
norm $\|R\|$)
 may be introduced as independent additional variables. 
To comply with the general canonical principle, their density of probability $p'(R)$ has to 
tend to a delta distribution (centered in $0$) at the thermodynamical limit. 
A suitable example is $p'(R)=(n\lambda/\pi)^n\exp(-n \lambda \|R\|^2)$ for some constant $\lambda$.
\bibitem{classinote}Note that this particular scheme, 
adapting to the case in question the optimal classical strategy for the transmission 
of coherent states \cite{braunstein00}, is not necessarily optimal in this instance but 
is feasible with current technology.
\end{thebibliography}
\end{document}